\documentclass[12pt,preprint]{aastex}

\shorttitle{Additional \ion{He}{2} Quasar Sightlines} 
\shortauthors{}

\begin{document}

\title{Ten More New Sightlines for the Study of Intergalactic Helium, and Hundreds of Far-UV-Bright Quasars, from SDSS, {\it GALEX}, and {\it HST}\altaffilmark{1}}

\author{David Syphers\altaffilmark{2},
Scott F. Anderson\altaffilmark{3},
Wei Zheng\altaffilmark{4},
Daryl Haggard\altaffilmark{3},
Avery Meiksin\altaffilmark{5},
Donald P. Schneider\altaffilmark{6},
Donald G. York\altaffilmark{7,8}
}

\altaffiltext{1}{Based on observations with the NASA/ESA Hubble Space Telescope obtained at the Space Telescope Science Institute, which is operated by the Association of Universities for Research in Astronomy, Incorporated, under NASA contract NAS5-26555.}

\altaffiltext{2}{Physics Department, University of Washington, Seattle, WA 98195; dsyphers@phys.washington.edu}

\altaffiltext{3}{Astronomy Department, University of Washington, Seattle, WA 98195; anderson@astro.washington.edu}

\altaffiltext{4}{Department of Physics and Astronomy, Johns Hopkins University, Baltimore, MD 21218; zheng@pha.jhu.edu}

\altaffiltext{5}{Scottish Universities Physics Alliance (SUPA), Institute for Astronomy, University of Edinburgh, Royal Observatory, Edinburgh EH9 3HJ, United Kingdom}

\altaffiltext{6}{Pennsylvania State University, Department of Physics \& Astronomy, 525 Davey Lab, University Park, PA 16802}

\altaffiltext{7}{Department of Astronomy and Astrophysics, The University of Chicago, 5640 South Ellis Avenue, Chicago, IL 60637}

\altaffiltext{8}{Enrico Fermi Institute, University of Chicago, 5640 South Ellis Avenue, Chicago, IL 60637}

\begin{abstract}

Absorption along quasar sightlines remains among the most sensitive direct measures of \ion{He}{2} reionization in much of the intergalactic medium (IGM).
Until recently, fewer than a half-dozen unobscured quasar sightlines suitable for the \ion{He}{2} Gunn-Peterson test were known; although these handful demonstrated great promise, the small sample size limited confidence in cosmological inferences.
We have recently added nine more such clean \ion{He}{2} quasars, exploiting SDSS quasar samples, broadband UV imaging from {\it GALEX}, and high-yield UV spectroscopic confirmations from {\it HST}.
Here we markedly expand this approach by cross-correlating SDSS DR7 and {\it GALEX} GR4+5 to catalog 428 SDSS and 165 other quasars with $z>2.78$ having likely ($\sim 70\%$) {\it GALEX} detections, suggesting they are bright into the far-UV. 
Reconnaissance {\it HST} Cycle 16 Supplemental prism data for 29 of these new quasar-{\it GALEX} matches spectroscopically confirm 17 as indeed far-UV bright.
At least 10 of these confirmations have clean sightlines all the way down to \ion{He}{2} Ly$\alpha$, substantially expanding the number of known clean \ion{He}{2} quasars, and reaffirming the order of magnitude enhanced efficiency of our selection technique.
Combined confirmations from this and our past programs yield more than twenty \ion{He}{2} quasars, quintupling the sample.
These provide substantial progress toward a sample of \ion{He}{2} quasar sightlines large enough, and spanning a sufficient redshift range, to enable statistical IGM studies that may avoid individual object peculiarity and sightline variance. 
Our expanded catalog of hundreds of high-likelihood far-UV-bright QSOs additionally will be useful for understanding the extreme-UV properties of the quasars themselves.

\end{abstract}


\section{Introduction}

The intergalactic medium (IGM) may be the largest repository of baryons in the Universe, perhaps with baryonic matter content exceeding that which collapsed into stars and galaxies \citep[e.g., see the recent review by][]{mei07}.
The reionization of the IGM's primordial hydrogen may have occurred as early as redshifts of about $z\sim11$ \citep[e.g., WMAP results of][]{dun09} to as late as $z\sim6-7$ 
\citep[as inferred from quasar hydrogen absorption line measures, e.g.,][]{fan06}.
The veil of the ``dark ages" was lifted by hydrogen reionization, kindled by the ultraviolet (UV) radiation of massive early stars and star-forming galaxies, with a later ionizing contribution from quasars.
While \ion{He}{1} was probably ionized in the same epoch as hydrogen due to similar ionization potentials, the later, harder-spectrum ionizing contribution from quasars was likely necessary for reionizing IGM \ion{He}{2}.

Much of the IGM in the present-day Universe is confirmed to be highly ionized, with a hydrogen neutral fraction $x_{\rm{H\: I}}=N_{\rm{H\: I}}/N_{\rm{H\: II}} \sim 10^{-5}$ \citep{fan06}.
A sensitive measure of neutral hydrogen is to examine quasar spectra for the saturated trough shortward of hydrogen Ly$\alpha$ caused by even a very modest amount of neutral IGM gas, i.e., the classic Gunn-Peterson effect \citep{GP65}. 
Hydrogen Gunn-Peterson studies have been expanded in the last few years to include around twenty high-redshift quasar sightlines at $z\sim6$ \citep{fan06}, mainly derived from the high quasar selection efficiency and large sky coverage of the Sloan Digital Sky Survey \citep[SDSS;][]{yor00}.
The first few $z\sim6$ hydrogen studies returned diverse results and interpretations, showing that---because of possible individual object and/or sightline variance, including IGM patchiness---multiple independent clean quasar sightlines were required in order to clarify the ensemble picture of hydrogen Gunn-Peterson measures \citep[e.g.][]{whi03,fan06,tot06,bol07,bec07}.

However, it was helium, the second major primordial element in the IGM, that provided the first detection of the Gunn-Peterson effect \citep[using {\it HST};][]{jak94}.
While hydrogen predominates, it is nearly completely ionized in the IGM by the era of numerous quasars at $z\sim 2$--$4$; at such redshifts, singly ionized helium is instead the strongest absorber in much of the IGM accessible to study via background quasars. 
At $z\sim 2$--$4$, \ion{He}{2} outnumbers \ion{H}{1} by a factor of $\eta \sim 50$--$100$ \citep{fec06}.
Because kinetic velocities evidently dominate over thermal velocities in the IGM \citep{zhe04a}, hydrogen and helium have similar Doppler parameters, and therefore \ion{He}{2} has a higher opacity than \ion{H}{1} by $\eta/4 > 10$.
The \ion{He}{2} Ly$\alpha$ transition thus provides the more sensitive measure of much of the highly ionized IGM, potentially allowing rich cosmological inferences about the epoch of \ion{He}{2} reionization (delayed versus hydrogen), characteristics of the ionizing background radiation, and estimates of the cosmic baryon density in the IGM.

Despite providing the breakthrough Gunn-Peterson detection \citep{jak94}, helium studies have historically been limited to only a very small handful of suitable quasar sightlines.
The paucity arises as \ion{He}{2} Ly$\alpha$ absorption occurs in the extreme UV (304 \AA\ rest-frame), observable from space in only about $3\%$ (by random) of high redshift quasars which lie by chance along sightlines with little foreground absorption.
Detailed UV spectroscopic studies at good signal-to-noise (S/N) and resolution of such historical ``clean \ion{He}{2} quasars" are limited to just 4 objects that also span the limited redshift range of $z=2.7$--$3.3$: HS~1700+6416 at $z=2.72$ \citep{dav96,fec06}, HE~2347-4342 at $z=2.88$ \citep{rei97,kri01,sme02,zhe04a,shu04}, PKS~1935-692 at $z=3.18$ \citep{and99}, and Q0302-003 at $z=3.29$ \citep{jak94,hog97,hea00}.
One additional quasar, QSO 1157+3143 at $z=3.0$ \citep{rei05}, was observed more recently in the UV at similarly high spectral resolution.

Studies of these few historical clean \ion{He}{2} quasars did verify their science potential.
They collectively reveal that the IGM \ion{He}{2} optical depth increases from $\tau\sim 1$ near $z=2.5$ to $\tau >4$ at $z=3.3$. 
This is consistent with theoretical perspectives that the \ion{He}{2} reionization epoch may have occurred near $\sim3$ \citep{sok02,pas07}, and indirect observational evidence from spectral hardening at $z \sim 3$ \citep{son96,aga05,aga07}.
A sharp change in the \ion{H}{1} Ly$\alpha$ forest opacity at $z \sim 3.2$ \citep{ber03,fau08} may be a further indication for \ion{He}{2} reionization \citep{the02}, although this feature is not universally seen \citep{kim02} and the interpretation is uncertain \citep{mcq09,bol09}. 
\ion{He}{2} Gunn-Peterson studies have also provided empirical measures of the flux and spectrum of the ionizing background radiation at $z \sim 3$ and the baryon density ($\Omega_g\sim0.01$) in the IGM \citep[e.g.,][]{hog97,zhe04a,tit07}.

Yet, possible systematics in such a small sample size of just five well-studied helium sightlines limit confidence in global conclusions. 
For example, none of these five revealed a damping absorption profile redward of \ion{He}{2} \citep{mir98}, a predicted signature of the \ion{He}{2} reionization epoch.
This leaves unanswered whether that epoch might occur at $z>3.3$, or instead suggests that this signature is difficult to discern in practice.
For example, such a signature might be suppressed by the presence of quasar proximity zones \citep{mad00}.
An intriguing redward absorption profile is seen in a recent low-resolution (but good S/N) prism UV {\it HST} spectrum of another newly suggested \ion{He}{2} quasar at $z=3.8$, but is not evident in a second at $z=3.5$ \citep{zhe08}.
With only two such high-redshift quasar spectra for \ion{He}{2} it was uncertain whether such an absorption feature might be common at high redshift, or reflects something unusual about this particular $z=3.8$ sightline. 
(Note that the strength and breadth of this feature is dramatically larger than the anticipated reionization epoch signature.)
The diversity found in the \ion{H}{1} Gunn-Peterson results \citep{fan06} also presaged the need for a similarly large sample of clean \ion{He}{2} quasars to average over systematics of individual object peculiarity, sightline variance, and IGM patchiness.
This motivated the SDSS and {\it HST} searches of \citet{zhe04b,zhe05}.

The advent of the {\it Galaxy Evolution Explorer} \citep[{\it GALEX},][]{mor07} all-sky UV survey provided an order of magnitude more efficient selection technique for establishing such a substantially larger sample of clean \ion{He}{2} quasars, spanning a broad redshift range.
Our cross-correlation of $z>2.8$ SDSS (and other) quasar catalogs with the broadband UV sources from {\it GALEX} yields far-UV-bright (observed frame) quasars with good confidence.
In their rest frames, these quasars are bright in the extreme-UV (EUV, $\lambda \lesssim 912$ \AA\ rest-frame), and this is how we will refer to them henceforth.
We have previously used this cross-correlation technique with SDSS DR6 \citep{ade08} and {\it GALEX} GR2+3, combined with Cycle 15-16 reconnaissance observations with the {\it HST} Advanced Camera for Surveys/Solar Blind Channel (ACS/SBC), to identify with high efficiency nine new clean \ion{He}{2} quasar sightlines \citep[][hereafter, Paper I]{syp09}.

This paper presents a significant expansion to our Paper I identifications of \ion{He}{2} and other EUV-bright quasars, doubling our previous samples and providing substantial lists from which to draw for detailed EUV quasar and IGM studies.
Rapid past-year progress was enabled by the combination of expanded sky coverage now available from SDSS DR7 \citep{aba09} and {\it GALEX} GR4+5, plus recent Cycle 16 Supplemental {\it HST} ACS/SBC reconnaissance UV prism verification spectra of many new clean quasar sightlines.
In section 2 of this paper we describe our expanded SDSS DR7 and {\it GALEX} GR4+5 cross-correlation catalog of 428 far-UV detections of SDSS quasars at $z>2.78$, plus an additional 165 (non-SDSS) quasars with {\it GALEX} detections.
Section 3 presents results of our {\it HST} Cycle 16 Supplemental UV prism spectra for 29 new candidate \ion{He}{2} quasars drawn from the SDSS/{\it GALEX} matches.
Among these, 17 are spectroscopically confirmed as new EUV-bright quasars, with at least 12 new quasar sightlines ($z=3.17$--$3.65$) confirmed to have flux down to the \ion{He}{2} break (``\ion{He}{2} break quasars''); 10 of these are clean sightlines to \ion{He}{2} Ly$\alpha$ with no substantial redward absorption (``clean \ion{He}{2} quasars'').
Our newly confirmed cases represent substantial progress in filling in sparsely populated redshift regimes, and our combined programs quintuple the clean \ion{He}{2} quasar sample size versus the historical sample discussed above.
Section 4 discusses the sightline diversity of our new \ion{He}{2} quasar prism spectra; but we also present initial ensemble characteristics of our combined sample of all EUV-bright quasars with ACS/SBC prism data, including rest-frame spectral stacks in several redshift bins spanning $z=3.1-3.9$ to average over peculiarities of individual objects.
(Detailed spectral analysis will follow in a subsequent paper).
Section 5 provides a brief summary and concluding remarks.

\section{Selection of EUV-Bright Quasars}

Our identification of new \ion{He}{2} quasar sightlines starts with the cross-correlation of quasars found in SDSS (and other catalogs) with the {\it GALEX} broadband catalog of UV sources.
The details of the cross-correlation steps are provided in Paper I, but selected aspects are also presented briefly in this section, along with the expanded catalogs of EUV-bright (rest-frame) quasars provided herein.

\subsection{Expanded SDSS Quasar and {\it GALEX} UV-Source Catalogs}

The large area 5-filter optical imaging photometry and follow-up multiobject optical spectroscopy of SDSS efficiently identify and confirm large numbers of quasars \citep[e.g.,][]{ric02,sch07}.
The SDSS Data Release 7 \citep[hereafter, DR7;][]{aba09} used here includes nearly $\sim$12,000~deg$^2$ of optical imaging, with spectroscopy encompassing about $\sim9400$~deg$^2$ of sky, especially centered near the north Galactic cap; the expanded spectroscopic coverage of SDSS DR7 reflects an increase of nearly 30\% over the DR6 coverage used in Paper I.
The SDSS data are obtained with a dedicated 2.5m telescope \citep{gun06} at Apache Point Observatory, New Mexico, equipped with a large-format mosaic camera that can image of order $10^2$~deg$^2$ per night in 5 filters \citep[$u,g,r,i,z$;][]{fuk96}, along with a multifiber spectrograph that simultaneously obtains 640 spectra within a 7~deg$^2$ field.
The imaging database is used to select objects for the SDSS spectroscopic survey, which includes spectrophotometry ($\lambda/\Delta\lambda\sim1800$) covering 
3800-9200\AA\ for $10^6$ galaxies, $10^5$ quasars, and $5 \times 10^5$ stars.
Technical details on SDSS hardware and software, and astrometric, photometric, and spectral data may be found in a variety of papers: e.g., \citet{gun98}, \citet{lup99}, \citet{hog01}, \citet{sto02}, \citet{smi02}, \citet{pie03}, and \citet{ive04}.

We are especially interested here in SDSS quasars with $z>2.78$, for which the \ion{He}{2} Ly$\alpha$ line at rest-frame 304 \AA\ would be redshifted to an observed wavelength $>$1150\AA. 
Our specific choice of $z>2.78$ is motivated, in part, by the {\it HST} Servicing Mission 4 (SM4) revitalization of quality UV spectroscopy; the refurbished STIS is anticipated to be useful down to $\sim$1150 \AA, and the Cosmic Origins Spectrograph (COS) may even still have good response at this wavelength.
At the time of proposing and planning for our new {\it HST} observations described in section 3, there was not yet a vetted DR7 quasar catalog, and so we instead used the DR5 quasar catalog of \citet{sch07} supplemented with a set of SDSS DR7-only $z>2.78$ quasar candidates whose optical spectra we confirmed by eye. 
Our combined SDSS DR7 list used here thereby includes 9462 quasars at $z>2.78$ (see Figure 1).
We also supplement this SDSS quasar list with the 12\textsuperscript{th} ed. V\'{e}ron-Cetty \& V\'{e}ron (VCV) quasar catalog \citep{ver06}, discussed further below.

{\it GALEX} provides an overlapping large-area UV imaging survey \citep{mor07}.
Although the broadband {\it GALEX} observations generally do not suffice to confirm UV flux all the way down to \ion{He}{2} Ly$\alpha$, they do allow efficient culling to the small subset of quasars most likely to have flux well into the EUV (rest-frame).
{\it GALEX} images cover both the FUV ($\sim$1350-1750\AA) and NUV ($\sim$1750-2800\AA) bands, with the all-sky survey (AIS) UV images extending to $m_{\mbox{\scriptsize{AB}}} \sim 21$, and the much smaller area medium and deep surveys (MIS and DIS) extending to $m_{\mbox{\scriptsize{AB}}} \sim 23$ and $m_{\mbox{\scriptsize{AB}}} \sim 25$, respectively.
The {\it GALEX} GR4 and GR5 catalogs (henceforth GR4+5) used here are complementary data sets processed with the same pipeline and constituting one catalog.
GR4+5 covers $\sim$25,000~deg$^2$ of sky, an 80\% increase over GR2+3 used in Paper I.
Among the 9462 SDSS $z>2.78$ quasars discussed above, {\it GALEX} GR4+5 provides UV imaging coverage for 8413 (5829 quasars from the vetted DR5 catalog, and 2584 more from DR7 only).
It is these 8413 SDSS quasars that are actually matched to {\it GALEX} GR4+5 UV catalogs, and considered further below.

\subsection{Expanded Cross-Correlation Catalogs of EUV-Bright Quasars from SDSS/{\em GALEX} and VCV/{\em GALEX}}

Following our Paper I approach, we cross-correlate the 8413 SDSS DR7 $z>2.78$ quasars that fall in the {\it GALEX} footprint with the GR4+5 UV-source catalogs, to obtain 434 SDSS-quasar/{GALEX}-UV positional coincidences matching within 3$''$.
We then visually examined the {\it GALEX} images to remove six objects which were artifacts or too close to artifacts to be certain of a UV detection.
This is proportionally only about a quarter the number of {\it GALEX} artifacts similarly removed in Paper I, a clear demonstration of the improvements of the GR4+5 pipeline over the GR2+3 version (in particular, masking of the edges of {\it GALEX} plates).
SDSS optical images and spectra were also examined by eye, to produce an inspection flag noting a few other possible problems (described below).

As in Paper I, cross-correlations of Monte Carlo shifted catalogs were used to empirically estimate the expected background random SDSS/{\it GALEX} coincidence match rate, and the signal appears to dominate over the noise for $r \lesssim 3''$.
We again find that $\sim70\%$ of our cataloged SDSS/{\it GALEX} matches are likely genuine associations.
Note that due to the faintness of the targets we have proportionally far more {\it GALEX} detections on MIS and DIS plates than would be expected just from the geometric survey coverage.
The higher UV source density on these longer {\it GALEX} exposures increases the false match rate somewhat, which is why we prefer a direct empirical method for estimating false matches, and, for example, why we have somewhat different match statistics than \citet{bud09}.
Our estimate of the false match rate includes both happenstance matches to unrelated real UV sources, as well as matches to spurious {\it GALEX} sources; {\it HST} follow-up discussed in section 3 suggests we have both.
(Even if the flux errors given in the {\it GALEX} catalog are assumed to be Gaussian, we estimate that $\approx$1\% of our apparent matches are simply matches to noise.
In practice, this estimate is almost surely too low.
It should also be noted that our {\it HST} reconnaissance spectra extend down to the \ion{He}{2} break and generally have very limited wavelength overlap with the {\it GALEX} NUV filter.
Only about a third of our matches have detected FUV flux in {\it GALEX} GR4+5, and thus even real optical-UV matches may be undetected in the far-UV {\it HST} spectra.)

The 428 SDSS DR7 quasars at $z>2.78$, associated at high ($\sim$70\%) confidence with {\it GALEX} GR4+5 UV sources (3$''$ match radius) are cataloged in Table \ref{tab:TargetListSDSS}.
Our catalog tabulates the following information.
{\textit{Columns 1-5}}---Names, positions (RA, dec; J2000), redshifts, and $z$-band PSF magnitudes are all from SDSS.
The redshift is from the automated SDSS pipeline, but we verified these (via by-eye examinations of the SDSS spectra) to be reasonably accurate in each case.
{\textit{Columns 6-7}}---The {\it GALEX} catalog provides fluxes in $\mu$Jy, which we have converted to erg~s$^{-1}$~cm$^{-2}$~\AA$^{-1}$ using the following effective wavelengths of the {\it GALEX} broadband filters: 2316\AA\ for NUV, and 1539\AA\ for FUV.
{\textit{Column 8}}---Inspection flag. This is a 4-bit integer flag, reflecting our by-eye inspection SDSS images and spectra.
If there was a galaxy or possible galaxy nearby in projection in SDSS images (typically within $10''$), or a very near blue object (typically within $5''$), this flag gets a value of 1.
A value of 2 indicates a probable Lyman-limit system (LLS) or damped hydrogen Ly$\alpha$ absorber (DLA) in the SDSS optical spectrum.
A value of 4 indicates a possible broad absorption line quasar (BALQSO, having \ion{C}{4} absorption with a continuous width in excess of 2000~km~s$^{-1}$), and a value of 8 is given to those objects that are borderline on our BALQSO criterion.
As discussed in Paper I, these inspection flag values are additive.
{\textit{Column 9}}---{\it HST} observation flag.
The observation flag is 0 if the object has not been observed with {\it HST}, 1 if it has been observed with FUV (to near 304 \AA\ rest) spectroscopy through {\it HST} Cycle 16 Supplemental, 2 if it has been observed less conclusively (NUV spectroscopy, UV imaging, or pre-COSTAR observation), and 3 if it is a planned target for {\it HST} Cycle 17 observation.

Our expanded catalog of 428 high-confidence EUV-bright candidate quasars presented here more than doubles our earlier similar catalog of Paper I. 
Again reassuringly, in blind fashion, this expanded catalog recovers all of the previously confirmed \ion{He}{2} break quasars in the SDSS/{\it GALEX} footprint with $z>2.78$, including: Q0302-003 \citep[$z=3.29$;][]{jak94}, HS~1157+3143 \citep[$z=2.99$;][]{rei05}, SDSS~J1614+4859 \citep[$z=3.80$;][]{zhe05}, SDSS~J2346-0016 ($z=3.51$) and SDSS~J1711+6052 ($z=3.83$) \citep[both][]{zhe08}, and the other \ion{He}{2} quasars we previously confirmed in Paper I (excluding two faint ones cataloged in GR1, but not found in GR2+3 or GR4+5).  
Follow-up {\it HST} reconnaissance UV spectra for a subset of 29 of our new EUV-bright candidates are described below in Section 3.

We similarly expanded our search for other (non-SDSS) EUV-bright quasars, by matching the quasar catalog of \citet{ver06} to the expanded {\it GALEX} GR4+5 sky coverage as well.
Among the 1123 $z>2.78$ quasars in \citet{ver06} that are not in SDSS DR7, 993 are in the {\it GALEX} GR4+5 footprint.
Using the same $3''$ match radius as above, among these 993 additional quasars there are (after discarding one probable {\it GALEX} artifact) 165 matches to {\it GALEX} GR4+5 sources; we refer to these as the VCV/{\it GALEX} quasars. 
(Note that in Paper I we used $3 \farcs 5$ for VCV/{\it GALEX} quasar matches, but $3''$ is now also favored for VCV quasars in the current expanded data sets.)
From our mock catalog offset Monte Carlos, we estimate $\sim$90\% of the VCV/{\it GALEX} associations are likely true matches.
The higher match rate (and lower false match estimate), compared to the SDSS/{\it GALEX} correlation, is plausibly due to the VCV sample being systematically at lower redshift than the SDSS sample, as discussed in Paper I.

Table~\ref{tab:TargetListVeron} presents the expanded catalog for these 165 additional (non-SDSS) candidate EUV-bright quasars.
The tabulated columns in this case are as follows. 
{\textit{Columns 1-5}}---Names, positions (RA, dec; J2000), redshifts, and magnitudes are all from \citet{ver06} and the specific references cited therein.
The optical magnitude provided is a V magnitude unless otherwise marked as: photographic (*), red (R), infrared (I), or photographic O-plates (O).
{\textit{Columns 6-7}}---The {\it GALEX} catalog provides fluxes in $\mu$Jy, which we have converted to erg~s$^{-1}$~cm$^{-2}$~\AA$^{-1}$ using the same transformation as described above.
{\textit{Column 8}}---{\it HST} observation flag (defined as in Table 1, above).

For these VCV/{\it GALEX} EUV-bright candidates, no inspection flag was readily feasible, given the difficulty of obtaining uniform images and spectra for this heterogeneous sample.
Related to this issue, we note that since we did not ourselves examine the optical spectra of these additional VCV quasars by eye, and because the VCV catalog cannot fully represent the uncertainty of the quasar identification or redshift determination, some additional caution may be appropriate when considering observations of quasars from our VCV/{\it GALEX} EUV-bright catalog.
On the other hand, it is again reassuring that in our expanded (about 47\% larger) VCV/{\it GALEX} catalog we recover in a blind fashion both additional previously confirmed clean \ion{He}{2} quasars that might be anticipated to be among our VCV/{\it GALEX} finds: PKS~1935-692 \citep[$z=3.18$,][]{and99} and HE~2347-4342 \citep[$z=2.88$,][]{rei97}.

The 428 SDSS and 165 other $z>2.78$ quasars cataloged in this section represent significantly expanded resource lists for EUV studies of quasars, approximately doubling the sample we cataloged in Paper I.
These hundreds of EUV-bright quasars may find use in a variety of quasar and IGM science applications from the study of EUV quasar SEDs, emission lines, and BALs, to the study of low-redshift HI and high-redshift \ion{He}{2} absorbers.
In the next section we report corresponding results of our recent and markedly expanded reconnaissance UV spectral observations from {\it HST} of SDSS/{\it GALEX} (plus a few VCV/{\it GALEX}) quasars, relevant to our own primary current aim of identifying a large sample of new unobscured sightlines potentially suitable for the study of IGM helium.

\section{Reconnaissance {\it HST} Spectra of New \ion{He}{2}, and Other EUV Bright, Quasars}

Under {\it HST} Cycle 16 Supplemental (GO program 11982), we were approved for reconnaissance UV ACS/SBC prism spectra observations of 40 candidate \ion{He}{2} quasar sightlines to assess their suitability for future detailed IGM \ion{He}{2} studies.
A total of 31 of our targets were actually observed before the close of the short Cycle 16 Supplemental observing window (which ended just before SM4, and effectively limited the RAs of our targets to the range $RA>10^h$).
We observed 29 new candidate \ion{He}{2} quasars (26 from SDSS, 3 from VCV) with brief UV ACS/SBC prism spectra to verify their potential suitability for future detailed \ion{He}{2} studies, along with two other cases in our catalog (SDSS~1614+4859 and OQ172, discussed further below) that were previously observed in the UV, but heretofore lacked good $S/N$ spectra.
Our original {\it HST} target list was based on {\it GALEX} GR4 (the small expansion of GR5 was not then available), SDSS DR7, and VCV 12\textsuperscript{th} ed.
When creating the initial {\it HST} target list we used the slightly more generous matching criterion that the quasar/UV-source offset be $<3 \farcs 2$, although in practice only one target had $r>3''$.

These Cycle 16 Supplemental prism observations are nearly identical in character to those discussed in Paper I for earlier \ion{He}{2} candidates.
Reconnaissance UV spectra (plus accompanying F150LP direct UV images that set the prism wavelength zero points) encompass two {\it HST} orbits each.
Total prism exposures times on each quasar are about 4~ksec, typically derived from coadding six individual prism exposures of about 500-800s each.
The prism PR130L, covering a wavelength range of at least 1250-1850 \AA, was used to obtain spectra near the \ion{He}{2} breaks, while still limiting backgrounds due to geocoronal Ly$\alpha$.
(For a few of the lowest or highest-redshift quasars we also cautiously use prism data just outside this wavelength range.)

Our Cycle 16 Supplemental {\it HST} observations reveal that 17 of the 29 new candidates, along with (not unexpectedly) both SDSS~1614+4859 and OQ172, indeed have detectable far-UV flux in the ACS spectral data; the 18 of these covered by SDSS are plotted as triangles in Figure 1.
Our {\it HST} Cycle 16 Supplemental reconnaissance results thus reaffirm the high yield of EUV-bright quasars obtainable from SDSS/{\it GALEX} cross-correlations for $r<3''$ offsets (also see Paper I).
(One further quasar, Q1051+5728, is detected at relatively low S/N in direct F150LP images when they are stacked, but is too faint to have a detectable spectrum.
For the purposes of this paper, we consider this a non-detection.\footnote{Of the 12 non-detections in {\it HST}, 10 had other UV sources in the ACS/SBC field of view.
Three of these are within a few arcseconds of the target quasar position, close enough that they are likely the UV source detected by {\it GALEX}, despite being barely resolved in SDSS images.})

The 19 EUV-bright confirmations from our Cycle 16 Supplemental {\it HST} prism reconnaissance program are listed in Table~\ref{tab:ACSObs}. The following information is tabulated (with the optical data in all but the last case from SDSS).
{\textit{Columns 1-4}}---Names, positions (RA, dec; J2000), and redshifts are mainly from SDSS.
The redshifts are (mainly) from the SDSS automated pipeline, verified as reasonably accurate in our own by-eye examination of the spectra.
{\textit{Columns 5-6}}---The {\textit{GALEX}} catalog provides fluxes in $\mu$Jy, which we have converted to erg~s$^{-1}$~cm$^{-2}$~\AA$^{-1}$ using the following effective wavelengths of the {\textit{GALEX}} broadband filters: 2316\AA\ for NUV, and 1539\AA\ for FUV.
{\textit{Columns 7-8}}---Exposure time and date the target was observed with {\it HST} ACS. The exposure times tabulated are those actually included in the final coadded spectrum. 
(This differs somewhat from the total {\it HST} exposure time in one case where we discarded particularly noisy or misaligned exposures, and two cases where there were telescope guiding problems).
{\textit{Column 9}}---Indicates whether or not there is a sharp, statistically significant flux break at a wavelength consistent with that expected for \ion{He}{2} Ly$\alpha$.
The 12 new quasars (plus SDSS~1614+4859) identified in the table as likely having a break at \ion{He}{2} conservatively are required to show a break of comparable width to the prism instrumental resolution (with allowance for a possible proximity zone).
There are two objects with such a flux break that do not have clean sightlines, however, and the absorption found redward of \ion{He}{2} Ly$\alpha$ makes those two less attractive for follow-up \ion{He}{2} study (as noted in the table, footnote b).
For objects marked as questionable, our data can neither confirm nor rule out a flux break at \ion{He}{2} Ly$\alpha$, either because of telescope pointing issues, very faint sources, or lowered flux near \ion{He}{2} due to redward absorption.

We reduced the ACS 2-D spectra of the 19 cases with ACS/{\it HST} prism confirmations using the aXe software, version $1.6$ \citep{kum08}; for further details on the reduction process, please consult Paper I.
For most quasars we coadded all the multiple individual 500s-800s spectra, extracted using optimal weighting \citep{hor86}.
For a single faint object---SDSS~1047+3250---we discarded a few individual prism exposures that were substantially more noisy than the average.
These discards are even fewer than in Paper I, which we attribute to both slightly longer direct UV image exposure times and the relative paucity of extremely faint detections in the current work.
Both these effects contribute to well-defined centers in the direct UV images, leading to adequate extraction alignment for nearly all the new prism exposures.

Extracted, coadded spectra of the 17 new EUV-bright quasars confirmed from {\it HST} are presented in Figures 2a, 2b, and 2c.
The relative flux calibration of our prism data is more secure than the absolute spectrophotometry.
Examination of the background subtracted by aXe indicates that it retains some residual structure of the spectrum signal, suggesting that too much flux is being subtracted (the contamination occurs because only local background subtraction is possible for the ACS/SBC prisms); this leads to a possible systematic $\sim 10$--$20$\% underestimation of the absolute flux in the spectra as presented.
The spectra of Figure 2 are displayed with wavelength bins of width equal to half of the (non-linear) instrumental resolution: for PR130L (with a two-pixel resolution element), $R$ ranges from 380 at 1250\AA\ to 40 at 1850\AA.
The dotted vertical line in each spectrum of Figure~2  depicts the expected observed wavelength of \ion{He}{2} IGM absorption for that quasar sightline.

Before turning to a discussion emphasizing ensemble characteristics, a few {\it HST} spectral targets warrant further comment.
First, the VCV/{\it GALEX} quasar Q1943-1502 at $z=3.3$, despite the appearance of its reconnaissance spectrum (Figure 2c), may be a good target for further helium IGM studies.
Our ACS spectrum for this object is unfortunately of very poor quality, due to {\it HST} pointing issues during this observation.
There are no data at all from one orbit, and the other orbit was pointed without the Fine Guidance Sensors.
This led to a drift of about 2-4$''$ during individual exposures, giving both poor wavelength calibration and very poor background subtraction (e.g., it is quite plausible that the flux is indeed zero below 1290\AA).
In addition, the spectra are normalized to the entire exposure time, giving the spectrum an artificially low apparent flux, since the region extracted was exposed only for a shorter time due to drift.
However, this object has the highest {\it GALEX} broadband flux levels of any of our previously unexplored targets, and has both NUV and FUV detections at $>14 \sigma$.
The actual UV continuum spectral flux is possibly about a factor of 5-10 higher than what is shown in Fig. 2c.
Its literature redshift is somewhat uncertain, with \citet{cra97} identifying only a single strong line as \ion{H}{1} Ly$\alpha$, and deriving $z=3.29$ from this.
They implicitly adopt a statistical uncertainty of at least $\sigma_z=0.05$, which we use to mark our UV spectrum, showing that the break we see is indeed consistent with \ion{He}{2} Ly$\alpha$.
This object is therefore tentatively another new \ion{He}{2} quasar, although we conservatively mark it as uncertain in Table 3 and include it only as EUV bright in subsequent discussion.

Additionally, Figure 3 shows our ACS/SBC prism spectra of SDSS~1614+4859 at $z=3.8$ and OQ172 at $z=3.5$. Both these quasars are recovered in our SDSS/{\it GALEX} catalogs of section 2 (and Paper I), and both have earlier but modest S/N {\it HST} UV spectra published in the literature.
We were motivated by the intriguing results of \citet{zhe08} in a similar redshift range of $z=3.5-3.8$ to obtain ACS/SBC prism UV spectra of these two quasars with improved S/N near \ion{He}{2} Ly$\alpha$. 
SDSS~1614+4859 was first identified by \citet{zhe05} as a new clean \ion{He}{2} quasar; our improved S/N ACS/SBC spectrum shows that it is indeed a very clean sightline, with a marked sharp break at the expected location of \ion{He}{2} Ly$\alpha$. 
OQ172 \citep{jak93,lyo95} was amongst the earliest {\it HST} targets considered and rejected as a Gunn-Peterson target; again our higher $S/N$ ACS/SBC spectrum (Figure 3) affirms earlier conclusions, with strong absorption (probably due to \ion{H}{1}) redward of the expected \ion{He}{2} break location. 
With the availability of these directly comparable ACS/SBC spectra, we are able to also include these two individually interesting cases, where suitable, in the following ensemble considerations (although, for example, exclude them from selection efficiency statistics).

Finally, we note that in the remaining discussions we have uniformly re-reduced all the similar ACS/SBC prism data available for a total of 33 EUV bright quasars; these prism spectra collectively come from \citet{zhe08}, Paper I, and this paper. 
These updated and uniform re-reductions also allow us further confidence in the helium character of one additional quasar, SDSS~1319+5202, from Paper I.
We now include this quasar as an additional \ion{He}{2} break confirmation (whereas in Paper I, we had labeled it as an uncertain case), because although it does have some redward absorption, there is a sharp, strong break at \ion{He}{2} Ly$\alpha$ evident in our current reductions.

\section{Discussion}

\subsection{Selection Efficiency for EUV-Bright Quasars}

The high success rate of our Cycle 16 Supplemental {\it HST} observations reported here in confirming new EUV-bright quasars ($17/29 \approx 59$\%) and in finding new \ion{He}{2} quasars ($12/29\approx 41$\% with \ion{He}{2} breaks, and  $10/29 \approx 34$\% very clean down to \ion{He}{2}) again demonstrates the value of our quasar/{\it GALEX} cross-correlation catalogs (see also Paper I).
Although our new {\it HST} targets were not randomly selected, the few-month visibility constraints of {\it HST} Cycle 16 Supplemental, combined with the large number of targets, means that the sample discussed herein is somewhat representative of our (SDSS/{\it GALEX}) catalog as a whole.
Although we preferred higher quality (absence of LLS or DLA in the SDSS spectrum, for example) targets for the {\it HST} observations, we also pushed to high redshift (in several cases $z>4$), and for targets at a variety of redshifts, since we seek a sample to characterize the redshift evolution of \ion{He}{2} absorption.
To some degree, these two selection effects tend to cancel one another.
It should also be noted that our {\it HST} reconnaissance targets were selected from the $z>3.1$ subset of the targets listed in Tables \ref{tab:TargetListSDSS} and \ref{tab:TargetListVeron}.
Lower redshift targets are more likely to be true matches, given their systematically higher flux and lower chance of intervening absorption systems, on average.
Thus our $\sim 60$\% prism spectra success rate (for $z>3.1$ UV-bright objects) is not incompatible with the predicted $\sim 70$\% rate (for $z>2.78$ objects) found in Section 2 by matching offset catalogs.

When considering the full reconnaissance sample, also including Paper I results, we have observed with {\it HST} a total of 39 new targets that also are in our current quasar/{\it GALEX} cross-correlation catalogs (Tables 1 and 2) at offsets $r<3''$.
Of these, 27 (69\%) are confirmed with {\it HST} as EUV bright, with at least 22 (56\%) having \ion{He}{2} Ly$\alpha$ breaks, and at least 19 (49\%) being clean \ion{He}{2} sightlines.
(Here we do not include the two previously considered UV-bright quasars, SDSS~1614+4859 and OQ172, discussed in section 3.
We also do not include quasars with $r>3''$, of which we have three: two of our \ion{He}{2} quasars confirmed in Paper I, and one undetected target from the Cycle 16 Supplemental program.
However, if included, they have little effect on the actual percentages.)

Using the 39 {\it HST}-observed targets with $r<3''$, we may also re-examine the efficacy of our chosen quasar/{\it GALEX} match radius in light of the actual results of our ACS/SBC prism observations.
As one would expect, those objects verified to be UV bright from {\it HST} tend to have smaller offsets between their SDSS and {\it GALEX} positions, but this effect is not large.
For example, for the $r<3''$ sample, geometrically 50\% of random points should be at $r<2.12''$.
In practice, that region contains 68\% of our non-detections, and 85\% of our detections.
The appearance that the two sets of offset distances are different, but not substantially so, is quantified by a Kolmogorov-Smirnov (KS) test.
This shows that the two sets of positional offsets have a 12\% chance of being drawn from the same distribution.
We thus verify our 3$''$ offset limit a reasonable compromise between contamination and inclusiveness (since EUV-bright $z>2.78$ quasars are still quite rare).

\subsection{Ensemble \ion{He}{2} Character of EUV-Bright Quasars}

For our own primary science interests related to ionized helium in the IGM, a major result of this paper is the end-to-end selection and spectral verification of 12 new \ion{He}{2} break quasars (Table 3; all are from SDSS/{\it GALEX}/{\it HST}).
We find clean sightlines, with no obvious absorption redward of \ion{He}{2} Ly$\alpha$, in 10 of these 12 quasars.\footnote{The high spectral sensitivity and success rate of the current survey and those of Paper I mean that in addition to finding many clean, high-quality sightlines appropriate for follow-up, we have also found a few objects that, despite having a flux break at \ion{He}{2} Ly$\alpha$, may be of relatively marginal quality for follow-up.
However, note that SDSSJ1341+0756, despite an intervening LLS, has residual continuum flux comparable with several of our fainter apparently clean sightlines.
We cannot rule out LLS in the region between {\it HST} and SDSS spectra, $\sim$1850--3800\AA\ (observed), and this may indeed explain the faintness of some of our objects.}
Our approximately 34\% selection efficiency for clean \ion{He}{2} quasars in the current paper is again an order of magnitude better than most previous searches (which typically had yields of order 3-6\%).  
Our improved S/N ACS/SBC prism observations also strongly affirm the \citet{zhe05} conclusion that SDSS~1614+4598 is an excellent, clean, helium quasar sightline to $z=3.8$. 
Our ACS/SBC prism spectra of SDSS~1614+4859 (this paper) and SDSS~1137+6237 (Paper I) provide further evidence that the strong absorption redward of \ion{He}{2} seen in the UV spectrum toward SDSS~1711+6052 (another quasar at similar redshift) is---as argued by \citet{zhe08} for other reasons---unlikely to be a universal (and unexpectedly strong) signature of IGM helium reionization.
The ten new clean \ion{He}{2} quasars selected and verified herein, together with the nine such confirmed \ion{He}{2} quasars from Paper I and the earlier SDSS/{\it HST} confirmations of \citet{zhe04b,zhe05,zhe08} quintuple the sample of clean sightlines suitable for IGM helium studies. 
These 21 clean \ion{He}{2} quasar sightlines represent significant progress toward a statistical sample for the study of IGM helium, spanning much of the redshift range during which \ion{He}{2} reionization is expected to take place (Figure 1).

With the beginnings of a statistical sample of \ion{He}{2} and other EUV-bright quasars, we perform preliminary stacks of our UV spectra in the quasar rest frame, to initially examine redshift evolution in three different redshift bins (Figure 4).
We use all 33 EUV-bright ACS/SBC quasar spectra from the current work, Paper I, and the three earlier \citet{zhe05,zhe08} cases, and normalize each by its continuum flux in their ACS/SBC prism spectra.
Using the ensemble of spectra, we then calculate the median flux value at each wavelength, and bin to the prism resolution of the highest-redshift (lowest-resolution) quasar in each stack.
We correct each individual-object prism spectrum for Galactic reddening before stacking, using the analytical formulae of \citet{car89} and the dust maps of \citet{sch98}.
(We do not attempt to correct for other---e.g., host galaxy---reddening, or aggregate low-column-density \ion{H}{1} Ly$\alpha$ forest absorption, although these effects could be considered in a statistical sense).
Although individual features such as proximity zones, possible EUV emission lines, and possible flux recoveries are important (and will be analyzed in a future paper), the median stacks provide representative ensemble spectra unbiased by a few high S/N cases or by the selection or rejection of particular objects one might consider in mean stacks.
The redshift bins here are described in relative terms as low redshift ($3.1 < z < 3.3$), moderate redshift ($3.3 \leq z < 3.6$), and high redshift ($3.6 \leq z < 4.1$); this choice of redshift bins is somewhat arbitrary, but allows stacking of 10-12 individual object prism spectra within each bin.

At the low ACS/SBC prism resolution, there is little obvious evidence for dramatic evolution of these binned/median spectra with redshift.
Although several of the input objects have noteworthy absorption redward of the \ion{He}{2} Ly$\alpha$ break, in many cases presumably due to low-redshift hydrogen, the median stack in each redshift bin has a clean break at \ion{He}{2} Ly$\alpha$.
However, it may be notable that the high-redshift objects individually are more likely than either the low or moderate-redshift objects to have strong absorption redward of \ion{He}{2} Ly$\alpha$; see \citet{zhe08} for a discussion of several ideas about the origin of such absorption.
Further analysis of these and other combinations of stacked \ion{He}{2} EUV prism spectra is ongoing.
However, a sensitive search for the predicted, more subtle, damped redward absorption profile of \ion{He}{2} reionization will likely require quality, higher resolution follow-up UV spectra.
Our combined sample of more than twenty \ion{He}{2} quasars, spanning a wide redshift range, provides a resource list for such studies, complementary to the (mainly) lower redshift cases emphasized in the studies of the historical \ion{He}{2} quasars.
Although none of our new cases are as UV bright as the two most luminous and lowest-redshift ($z<2.9$) historical \ion{He}{2} quasars, HS~1700+6416 and HE~2347-4342, some do have far-UV flux comparable to the other two historically well-studied \ion{He}{2} quasars, Q0302-003 and PKS~1935-692.
Moreover, most of our new clean \ion{He}{2} quasars have far-UV flux within an order of magnitude of Q0302-003 and PKS~1935-692 (both at $z=3.2$--$3.3$), assuring that quality UV spectra are feasible in reasonable exposure times with the current generation of UV instruments (COS or STIS on the refurbished {\it HST}).
The redshift span of our \ion{He}{2} quasars will allow follow-up study spanning most of the reionization epoch.

\subsection{Other Ensemble Spectral Results for EUV-Bright Quasars}

In closing the discussion section, we note that the far UV bright quasars verified in {\it HST} reconnaissance may also find use in other ultraviolet studies, from low-redshift HI absorption to EUV quasar emission lines and SEDs.
For example, several EUV-bright objects discussed herein appear to have intervening hydrogen absorption systems that, although optically thick, do not completely block all UV flux; these may be of interest for the study of low-redshift LLS. 
Such low-redshift \ion{H}{1} LLS are likely found in the following quasars:
{\it 1026+3531 (z=3.65)}---This object has a clear \ion{He}{2} Ly$\alpha$ break, but apparently also absorption due to a LLS is at $z \sim 0.9$.
This LLS has a column density of roughly $N_{\rm{H\:I}} \sim 6 \times 10^{17}$~cm$^{-2}$, based on an optical depth of $\tau_{LL} \approx 3.74^{+0.56}_{-0.32}$.
(Uncertainties quoted for the optical depths in this subsection are 68\% confidence interval limits, and incorporate only random errors.)
The flux below the hydrogen Lyman continuum break is estimated near that break, avoiding possible emission lines and the rise near \ion{He}{2} Ly$\alpha$.
The $\nu^3$ flux recovery leads to a factor of $\sim 6$ recovery by the \ion{He}{2} Ly$\alpha$ break.
{\it 1341+0756 (z=3.30)}---This object has a clear \ion{He}{2} Ly$\alpha$ break, but also a dramatic break redward, perhaps due to hydrogen continuum absorption. 
This possible LLS is at $z \sim 0.9$, and has a column density of roughly $N_{\rm{H\:I}} \sim 5 \times 10^{17}$~cm$^{-2}$, based on an optical depth of $\tau_{LL} \approx 3.04^{+0.20}_{-0.15}$.
The $\nu^3$ flux recovery leads to a factor of $\sim 5.6$ recovery by the \ion{He}{2} Ly$\alpha$ break.
{\it 1445+0958 (z=3.54)}---This object is OQ172 discussed in section 3, and targeted by \citet{lyo95} with FOS and found to have a strong flux break $\sim$100\AA\ (observed) redward of \ion{He}{2} Ly$\alpha$.
Our higher S/N spectrum (Fig. 3) suggests that the flux break is very extended, and the spectrum is consistent with having residual flux down to \ion{He}{2} Ly$\alpha$ (the final break occurs a single pixel from the expected \ion{He}{2} Ly$\alpha$ position, and thus within instrumental resolution).
We concur with \citet{lyo95} that the large flux break is due to a hydrogen LLS at $z \sim 0.6$, finding a tentative column density of $N_{\rm{H\:I}} \sim 7 \times 10^{17}$~cm$^{-2}$, based on an optical depth of $\tau_{LL} \approx 4.24^{+0.25}_{-0.18}$.
We note that \citet{lyo95} found no sign of \ion{H}{1} Ly$\alpha$ absorption around $\lambda \sim 1970$\AA, but their UV sensitivity was low enough to leave this consistent with our column density.
With a single-pixel resolution of $\sim$30\AA\ and extremely low sensitivity, our data can add no constraint at 1970\AA.

In addition to absorption studies, two other aspects of our spectral stacks worth considering are related to EUV quasar emission line and continuum radiation; we also present some initial comments on these aspects as well.
It is clear empirically (though also known theoretically) that \ion{He}{2} Ly$\alpha$ emission does not dominate the quasar spectrum the way \ion{H}{1} Ly$\alpha$ does---indeed, it scarcely seems present in the median stacks of Figure 4, though there appear to be individual exceptions to this rule (e.g., see SDSS~1442+0920 in Paper I).
This result was empirically also seen in the better studied \ion{He}{2} quasars with higher S/N spectra \citep{and99,hea00,sme02,fec06,zhe08}.
(QSO 1157+3143 \citep{rei05} is at $z=3.0$, which means any \ion{He}{2} Ly$\alpha$ from the quasar lies in a wavelength region difficult to observe due to geocoronal \ion{H}{1} Ly$\alpha$.)
There are similarly no other very strong EUV emission lines definitively present in our initial ensemble spectra.
However, the ACS/SBC (low and non-linear) prism resolution is not optimal for searches of subtle lines and blends, and individual objects again show some spectral diversity.

For constraints on the EUV (rest-frame) continua, we construct an ensemble spectral stack from our ACS/SBC prism data, similar to Figure 4 but now considering just the 13 objects with clean sightlines (lacking any obvious hydrogen absorption in our ACS prism spectra) and with $z \leq 3.5$ (to maximize the range of continuum wavelengths sampled).
For this preliminary result, we do not correct for aggregate low-column-density \ion{H}{1} Ly$\alpha$ forest absorption.
The ensemble continuum stack we assemble for these 13 quasars extends from 320\AA\ to 410\AA\ (rest-frame).
The upper limit comes from the redshift and instrument constraints, while the lower limit is chosen to avoid what appears to be a flux rise near \ion{He}{2} Ly$\alpha$, which is conceivably due to blended broad emission lines.
From our reddening-corrected ensemble median stack, we derive a preliminary EUV slope of $\alpha = -1.57 \pm 0.45$, for $f_{\nu} \sim \nu^{\alpha}$, which agrees with the EUV quasar slope found by \citet{tel02}, who looked primarily at somewhat longer wavelengths.
(Neither of these results are consistent with the harder EUV slope found by \citet{sco04} using FUSE to examine lower luminosity, lower redshift quasars.)
Note that there is considerable diversity among individual objects, with individual EUV slopes ranging from $\alpha = 3.69 \pm 0.78$ to $\alpha = -4.22 \pm 0.26$.
Large diversity in slopes is also seen at longer UV wavelengths in optical (observed-frame) spectroscopy \citep[e.g.,][]{dav07}, although the range we find may be even greater, perhaps due to a greater sensitivity of the EUV region to intrinsic or host-galaxy reddening.
(SMC-like extinction, which might describe quasar host galaxies better than Galactic extinction curves \citep{hop04}, rises even more sharply in the far-UV than does Galactic extinction \citep{pei92}.)
In any case, such direct ensemble constraints on the quasar EUV slope may find use in modelling \ion{He}{2} IGM reionization, as well as radiation emission and absorption processes internal to quasars themselves.

\section{Summary and Conclusions}

Utilizing recently expanded sky coverage from SDSS DR7 and {\it GALEX} GR4+5, we provide an expanded catalog of 428 SDSS quasars at $z>2.78$ that are likely ($\sim 70\%$) associated with {\it GALEX} ultraviolet sources, and thus bright in the EUV (rest-frame).
These likely EUV-bright quasars may find potential use for a variety of IGM and quasar studies, from EUV quasar SED continua, line emission, and intrinsic absorption lines, to probes of intervening \ion{H}{1} at low redshift, as well as the authors' own principal interest in the \ion{He}{2} Gunn-Peterson test for the IGM. 
Our EUV-bright quasar/{\it GALEX} catalogs are two orders of magnitude larger than most previous similar dedicated efforts, and double the size of our own previous lists from Paper I.

From our {\it HST} Cycle 16 Supplemental program, we herein also report new reconnaissance UV spectral observations with the ACS/SBC prisms for 31 quasars, selected from our optical/UV catalogs to encompass a wide redshift range. 
These {\it HST} observations spectroscopically verify that about 60\% are indeed EUV bright.
At least 12 new quasars are herein confirmed in our {\it HST} prism spectra as having \ion{He}{2} breaks, with 10 new very clean sightlines suitable for \ion{He}{2} Gunn-Peterson studies.
One previously UV-detected quasar at $z=3.8$ \citep{zhe05} is confirmed at good S/N to be free of any strong absorption all the way down to its clean \ion{He}{2} Ly$\alpha$ break.
Our SDSS/{\it GALEX}/{\it HST}-reconnaissance selection efficiency for clean sightlines ($\sim 34$\%) is an order of magnitude better than most previous similar \ion{He}{2} quasar searches; along with the earlier Zheng et al. SDSS/{\it HST} confirmations, our combined programs approximately quintuple the number of spectroscopically confirmed \ion{He}{2} quasars.
This set of new confirmed \ion{He}{2} sightlines provides a significant advance toward a sample statistically large enough to average over individual object peculiarity and sightline variance, although additional confirmations are still desirable (e.g., in the higher redshift bins).

We also present preliminary median stacks of our UV {\it HST} quasar spectra in three redshift bins.
These reveal no dramatic evolution of ensemble \ion{He}{2} absorption spectra with redshift, though our low-resolution prism stacks are not optimal for sensitive constraints on subtle features.
The median EUV continuum slope is consistent with earlier estimates for high-redshift quasars.
Analysis of the ensemble redshift evolution of \ion{He}{2}, and detailed analyses of individual \ion{He}{2} quasar spectra, would benefit from higher-resolution quality UV follow-up with instruments such as COS and STIS on the refurbished {\it HST}.
Our sample of new \ion{He}{2} quasar sightlines, spanning a substantial redshift range, provides an excellent resource list for such future investigations.
Most of these objects are bright enough for good-quality follow-up, and although none are as bright as the very luminous $z<3$ HS~1700+6416 or HE~2347-4342, they allow study of the interesting $z=3$--$4$ region, where the vast majority of \ion{He}{2} reionization took place.

In closing we note that in contrast to hydrogen reionization, helium reionization may be expected to show very strong fluctuations, as the reionization source for helium is thought to be rare, bright quasars. 
Some models predict that this will lead to helium transmission windows present in the IGM even during the early stages of helium reionization \citep{fur08}, as opposed to the hydrogen case where Gunn-Peterson troughs are primarily sensitive to the very latest stages of reionization.
Apart from these fluctuations in ionizing flux, theoretical work shows that IGM density fluctuations allow \ion{He}{2} Gunn-Peterson troughs to constrain neutral fractions much higher than experience with hydrogen would otherwise suggest, to $x_{\rm{He\:II}}\sim1$\% for \ion{He}{2} Ly$\alpha$ and $\sim10$\% for higher order lines, as opposed to $x_{\rm{H\:I}} \sim 0.001$\% for hydrogen \citep{mcq09a}.
As \ion{He}{2} reionization is theoretically predicted to occur in the redshift range $3 \lesssim z \lesssim 4$ \citep{sok02}, with overlap of \ion{He}{3} regions possibly occurring as early as $z \sim 4$ \citep{wyi03}, higher redshift \ion{He}{2} quasars ($z \gtrsim 3.2$-3.5), such as those identified here, should prove useful in providing long sightlines to study the evolution of the IGM as reionization progresses.
The expected strong helium IGM fluctuations also affirm that large numbers of sightlines are necessary for confident statistical conclusions, an expectation strengthened by the diversity we find in individual EUV spectra for our more than twenty \ion{He}{2} quasar sightlines explored in reconnaissance.
 
\acknowledgments

We gratefully acknowledge support from NASA/{\it GALEX} Guest Investigator grants NNG06GD03G and NNX09AF91G.
Support for {\it HST} Program numbers 10907, 11215, and 11982 was provided by NASA through grants from the Space Telescope Science Institute, which is operated by the Association of Universities for Research in Astronomy, Incorporated, under NASA contract NAS5-26555.

Funding for the SDSS and SDSS-II has been provided by the Alfred P. Sloan Foundation, the Participating Institutions, the National Science Foundation, the U.S. Department of Energy, the National Aeronautics and Space Administration, the Japanese Monbukagakusho, the Max Planck Society, and the Higher Education Funding Council for England. The SDSS Web Site is http://www.sdss.org/.

The SDSS is managed by the Astrophysical Research Consortium for the Participating Institutions. The Participating Institutions are the American Museum of Natural History, Astrophysical Institute Potsdam, University of Basel, University of Cambridge, Case Western Reserve University, University of Chicago, Drexel University, Fermilab, the Institute for Advanced Study, the Japan Participation Group, Johns Hopkins University, the Joint Institute for Nuclear Astrophysics, the Kavli Institute for Particle Astrophysics and Cosmology, the Korean Scientist Group, the Chinese Academy of Sciences (LAMOST), Los Alamos National Laboratory, the Max-Planck-Institute for Astronomy (MPIA), the Max-Planck-Institute for Astrophysics (MPA), New Mexico State University, Ohio State University, University of Pittsburgh, University of Portsmouth, Princeton University, the United States Naval Observatory, and the University of Washington.

\begin{figure}
\figurenum{1}
\epsscale{1.0}
\plotone{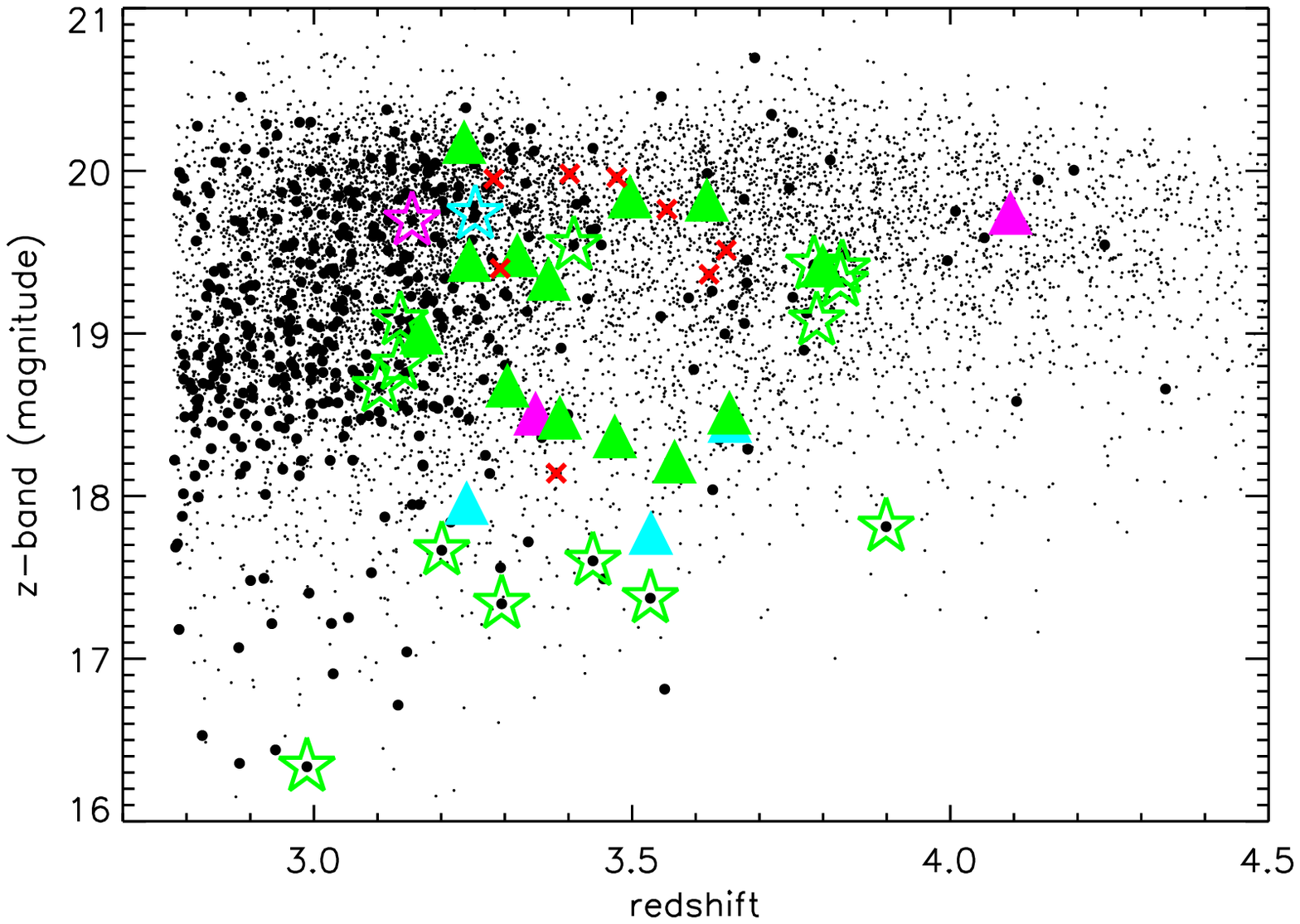}
\caption{A portion of the redshift-magnitude diagram for SDSS DR7 quasars (small black points).
Larger black circles indicate the 428 matches to {\it GALEX} GR4+5 UV sources discussed in section 2.
Solid triangles indicate the 18 SDSS quasars confirmed in the current paper to be far-UV bright, while open stars indicate observations from the literature (including Paper I).
The symbols have the following color coding: green indicates those objects with a statistically significant flux break at \ion{He}{2} Ly$\alpha$ (including two that also have strong redward absorption, as noted in Table 3), cyan those for which we can neither clearly detect such a flux break nor strongly rule it out (due to strong redward absorption or various calibration uncertainties), and magenta those whose flux clearly breaks to statistically zero at a wavelength redward of \ion{He}{2} Ly$\alpha$ (due to intervening absorption).
Small red x's indicate objects in our SDSS/{\it GALEX} catalog targeted but not detected by ACS/SBC in either the program of the current paper or those of Paper I.
Three additional SDSS/{\it GALEX} matches (not shown here) lie at $z=4.5$--$5.1$.
We obtained {\it HST} prism observations for two of these, but neither is confirmed as far-UV bright.}
\end{figure}

\clearpage

\begin{figure}
\figurenum{2a}
\epsscale{1.0}
\plotone{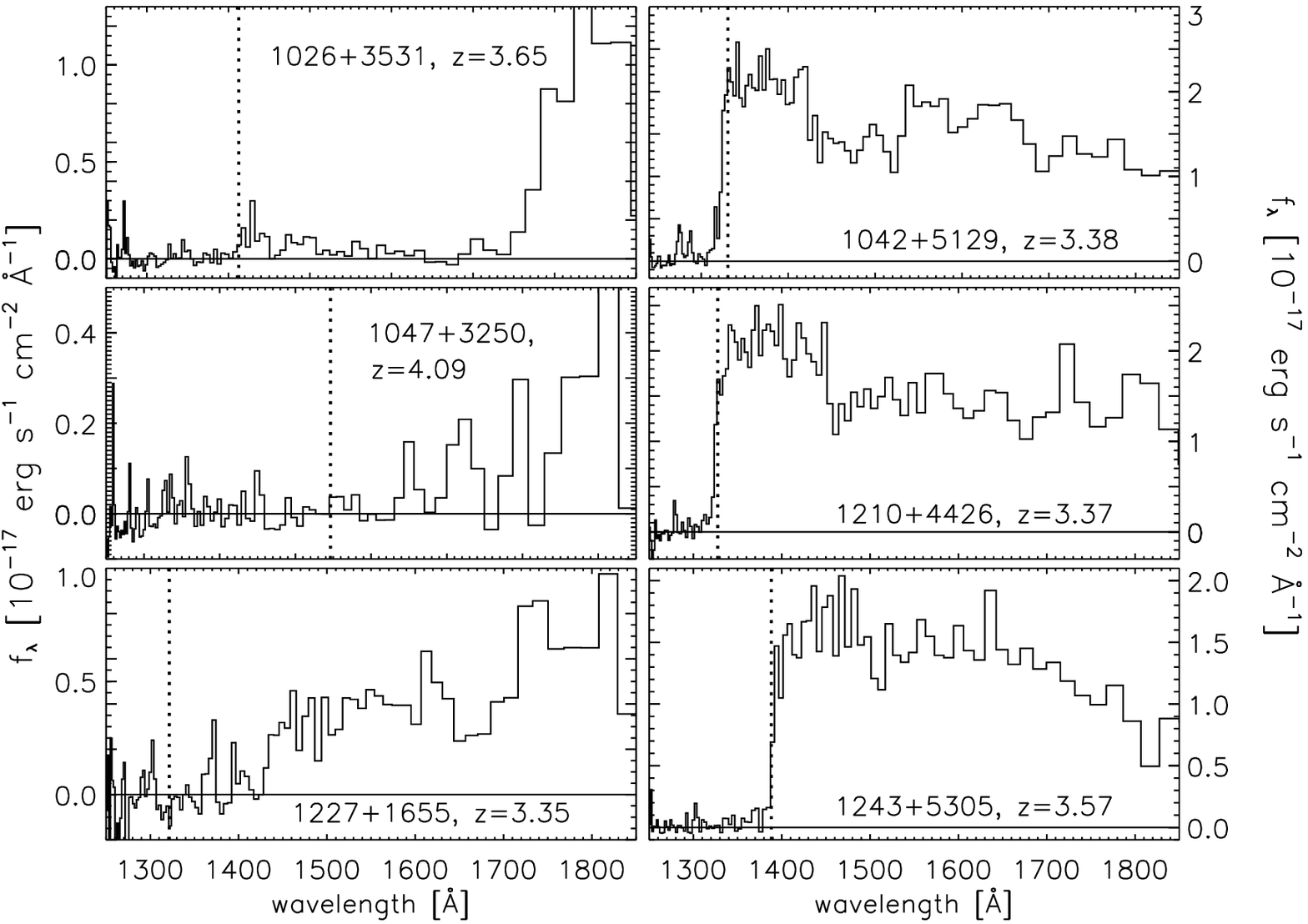}
\caption{ACS spectra for the first 6 new far-UV bright quasar discoveries listed in Table \ref{tab:ACSObs}.
The vertical dotted line shows the anticipated position of \ion{He}{2} Ly$\alpha$ for the quasar redshift.
The horizontal solid line indicates zero flux.
Spectra are displayed such that one spectral resolution element corresponds to two plotted bins.}
\end{figure}

\clearpage

\begin{figure}
\figurenum{2b}
\epsscale{1.0}
\plotone{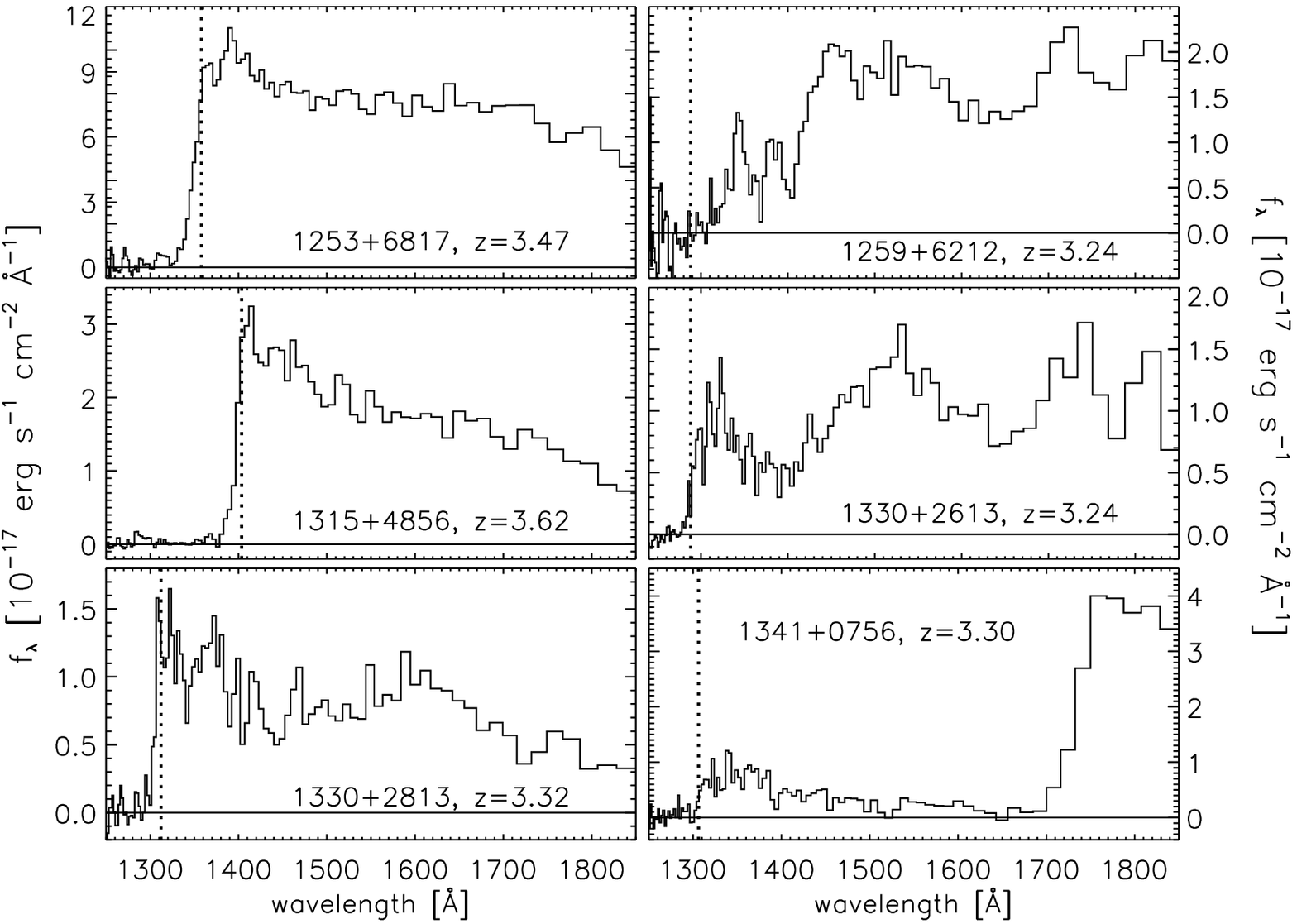}
\caption{ACS spectra for the second 6 new far UV-bright quasar discoveries listed in Table \ref{tab:ACSObs}.
The vertical dotted line shows the anticipated position of \ion{He}{2} Ly$\alpha$ for the quasar redshift.
The horizontal solid line indicates zero flux.
Spectra are displayed such that one spectral resolution element corresponds to two plotted bins.}
\end{figure}

\clearpage

\begin{figure}
\figurenum{2c}
\epsscale{1.0}
\plotone{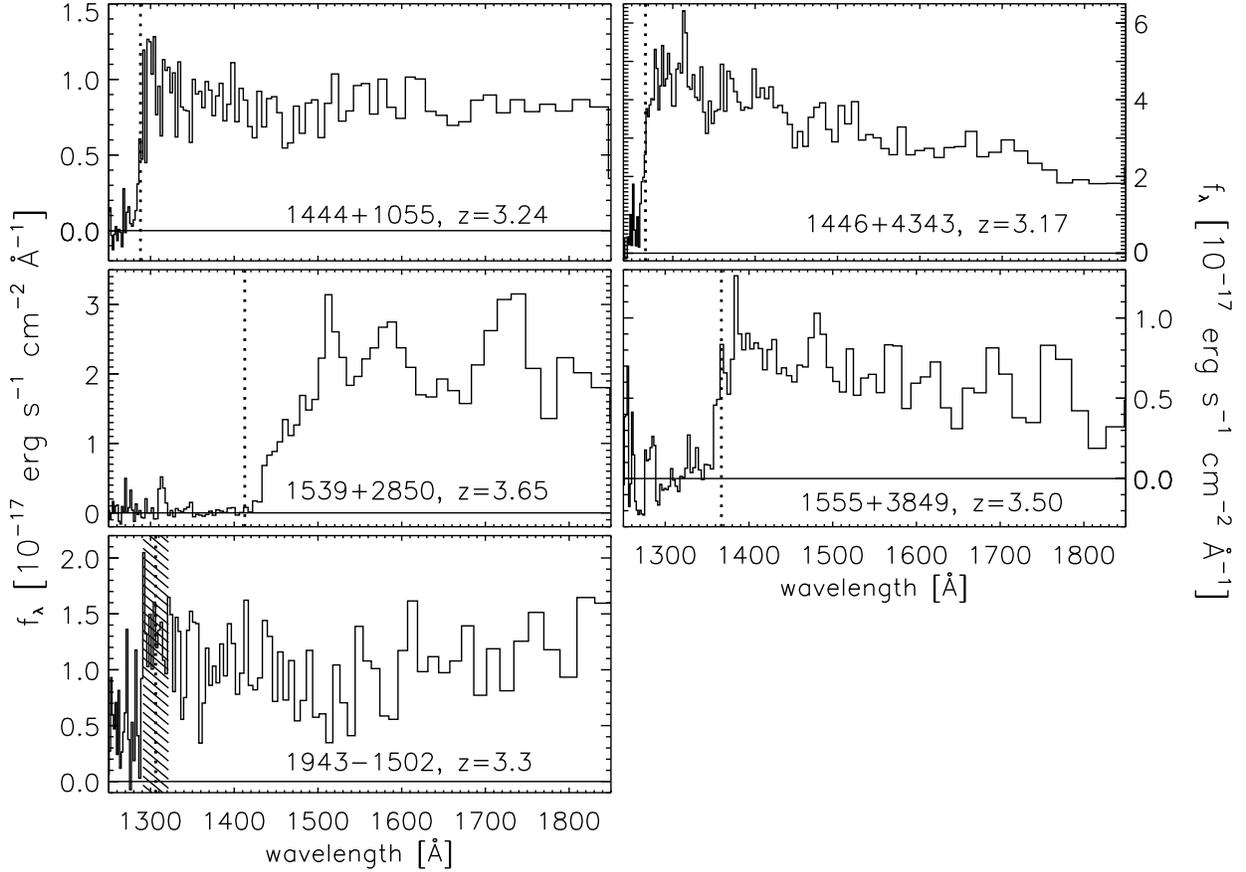}
\caption{ACS spectra for the last 5 new far UV-bright quasar discoveries listed in Table \ref{tab:ACSObs}.
The vertical dotted line shows the anticipated position of \ion{He}{2} Ly$\alpha$ for the quasar redshift.
The horizontal solid line indicates zero flux.
Spectra are displayed such that one spectral resolution element corresponds to two plotted bins.
For Q1943-1502: the shaded region of the plot shows the rough statistical uncertainty in the optical redshift; this UV spectrum has poor background subtraction, and the continuum flux is systematically underestimated, as discussed in section 3.}
\end{figure}

\clearpage

\begin{figure}
\figurenum{3}
\epsscale{0.6}
\plotone{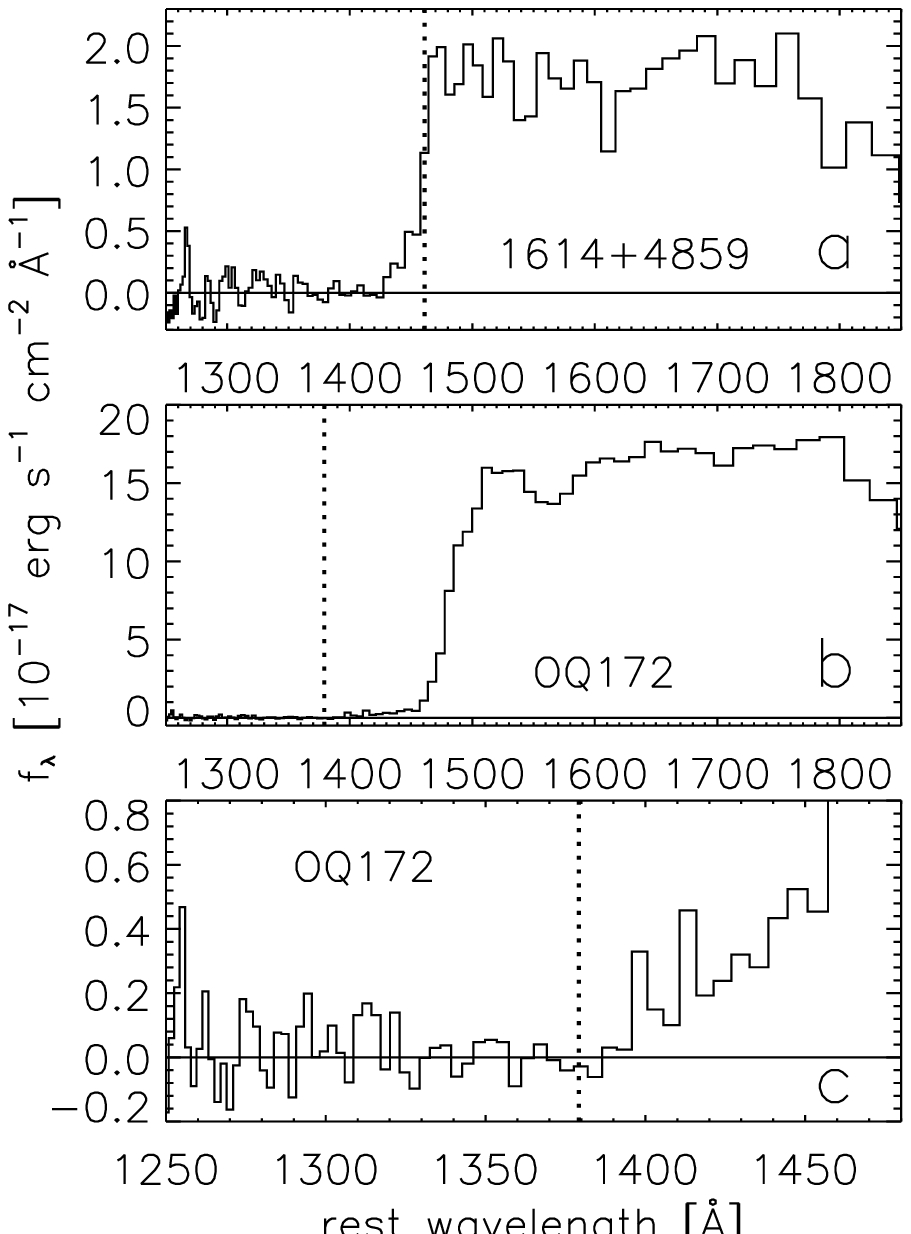}
\caption{ACS spectra for two previously considered HeII quasar candidates observed here at higher S/N.
The vertical dotted line shows the anticipated position of \ion{He}{2} Ly$\alpha$ for the quasar redshift.
The horizontal solid line indicates zero flux.
Spectra are displayed with two plotted bins per spectral resolution element.
(a) 1614+4859 ($z=3.81$), formerly observed at lower S/N by \citet{zhe05}, and here strongly verified to be a very clean \ion{He}{2} sightline to high redshift.
(b) 1445+0958, OQ172 ($z=3.54$), observed with {\it HST}/FOS by \citet{lyo95}, who noted the strong flux break strongly reaffirmed here to likely be due to low-redshift \ion{H}{1}.
OQ172 was also observed with {\it HST}/FOC by \citet{jak93}, who detected the continuum but found it unfeasible to obtain trustworthy data below 1550\AA\ (observed).
(c) A closer view of our OQ172 spectrum, which shows that although the intervening (\ion{H}{1}) system is optically thick, the more sensitive ACS/SBC prism can detect flux down to very near \ion{He}{2} Ly$\alpha$.}
\end{figure}

\clearpage

\begin{figure}
\figurenum{4}
\epsscale{1.0}
\plotone{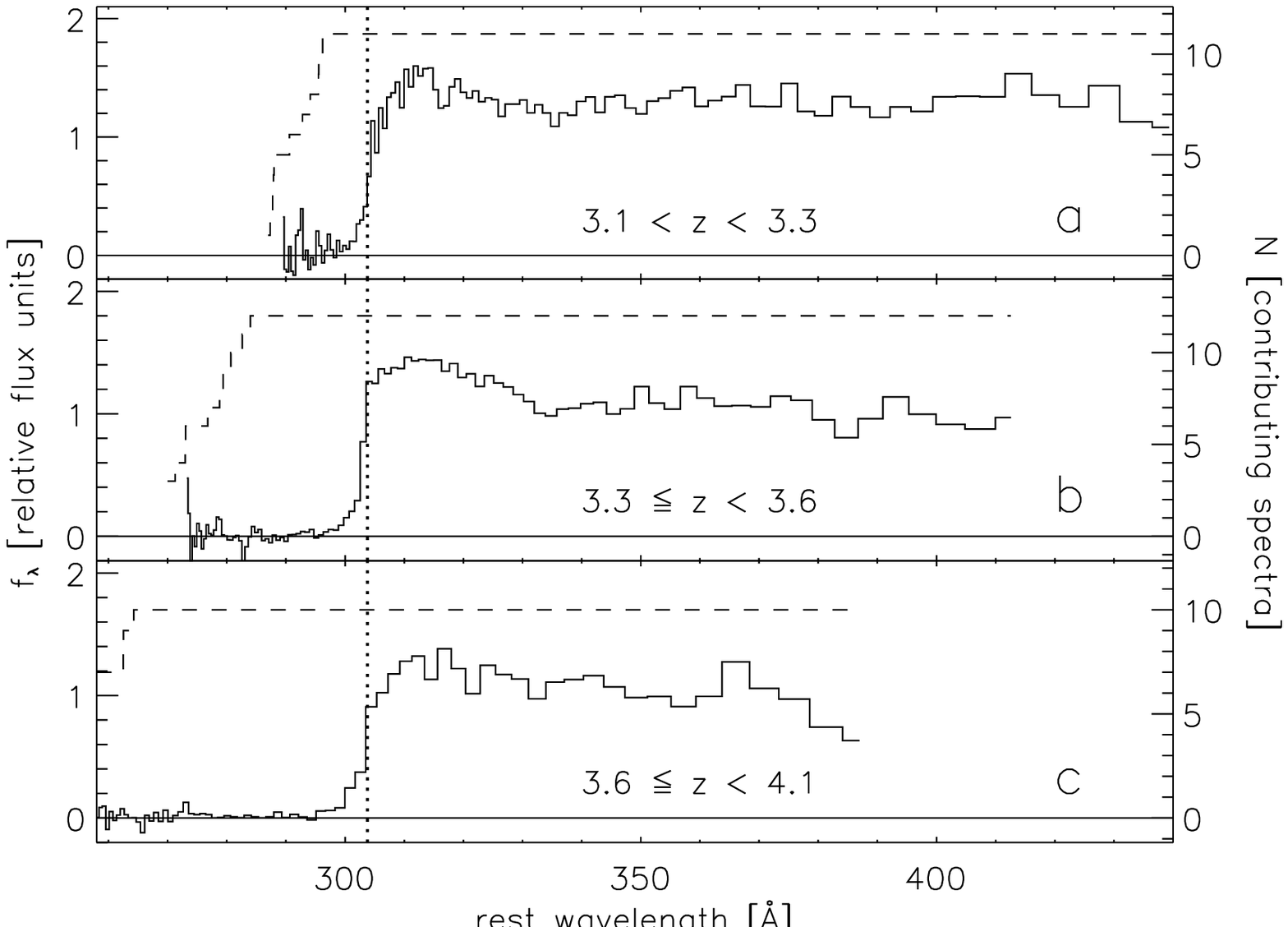}
\caption{Median rest-frame stacks of 33 EUV-bright quasar spectra observed with ACS/SBC prisms, including the 19 quasars of Table 3 in this paper, 12 EUV bright {\it HST} detections from Paper I, and two cases from \citet{zhe08}.
The solid line is the median quasar spectrum, the dashed line indicates the number of coadded spectra used in the stack at each wavelength, the vertical dotted line indicates the location of He II Ly$\alpha$, and the horizontal solid line indicates zero flux.
Spectra are binned to the resolution of the highest redshift (lowest resolution) quasar in the stack.
(a) Low redshift median stack; 11 objects with $3.1 < z < 3.3$.
(b) Moderate redshift median stack; 12 objects with $3.3 \leq z < 3.6$.
(c) High redshift median stack; 10 objects with $3.6 \leq z < 4.1$.}
\end{figure}

\clearpage


\begin{deluxetable}{lrrcccccc}
\tabletypesize{\footnotesize}
\rotate
\tablewidth{550pt}

\tablecaption{Catalog of Candidate EUV-Bright SDSS/{\it GALEX} Quasars 
\label{tab:TargetListSDSS}}

\tablehead{\colhead{Name} & \colhead{RA} & \colhead{Dec} & 
\colhead{Redshift} & \colhead{z-band} & \colhead{NUV flux} & \colhead{FUV flux} & \colhead{Insp.\tablenotemark{b}} & \colhead{Obs.\tablenotemark{c}} \\ 
\colhead{} & \colhead{(J2000)} & \colhead{(J2000)} & \colhead{} & \colhead{(mag)} & \colhead{($10^{-17}$)\tablenotemark{a}} & \colhead{($10^{-17}$)\tablenotemark{a}} & \colhead{} & \colhead{} } 
\startdata
SDSSJ000303.35-105150.7 & 0.763943 & -10.864079 & 3.65 & 19.00 & 0.69 & 3.19 & 10 & 0 \\
SDSSJ000316.39-000732.4 & 0.818289 & -0.125675 & 3.18 & 19.99 & 3.61 & \nodata & 0 & 0 \\
SDSSJ001132.02-000541.6 & 2.883429 & -0.094896 & 2.89 & 19.68 & 0.83 & \nodata & 0 & 0 \\
SDSSJ001626.54+003632.4 & 4.110588 & 0.609013 & 3.23 & 20.10 & 2.03 & \nodata & 2 & 0 \\
SDSSJ002223.94+355717.9 & 5.599748 & 35.954968 & 3.09 & \nodata & 2.52 & 8.71 & 0 & 0 \\
SDSSJ003017.11+005358.9 & 7.571309 & 0.899689 & 2.83 & 19.90 & \nodata & 0.73 & 0 & 0 \\
SDSSJ003420.62-010917.3 & 8.585921 & -1.154819 & 2.85 & 20.05 & 1.73 & \nodata & 0 & 0 \\
SDSSJ003828.89-010216.0 & 9.620364 & -1.037765 & 3.43 & 19.21 & 0.54 & \nodata & 0 & 0 \\
SDSSJ003939.96+152720.5 & 9.916517 & 15.455681 & 2.86 & 18.97 & \nodata & 2.80 & 0 & 0 \\
SDSSJ004323.43-001552.6 & 10.847631 & -0.264620 & 2.81 & 18.12 & 8.98 & \nodata & 8 & 0 \\
SDSSJ005215.65+003120.5 & 13.065205 & 0.522370 & 2.85 & 19.08 & 3.29 & 5.77 & 7 & 0 \\
SDSSJ005401.48+002847.8 & 13.506173 & 0.479941 & 3.41 & 19.80 & 3.91 & 6.93 & 3 & 0 \\
SDSSJ005653.25-094121.8 & 14.221886 & -9.689381 & 3.25 & 19.73 & 3.50 & \nodata & 0 & 1
\enddata

\tablecomments{This table is available in its entirety in a machine-readable form in the online journal. A portion is shown here for guidance regarding its form and content.}

\tablenotetext{a}{erg s$^{-1}$ cm$^{-2}$ \AA$^{-1}$, from {\textit{GALEX}}}
\tablenotetext{b}{The numeric inspection flag denotes possible problems with the object. The individual values are: $1$ for a very near neighbor in the SDSS 
image, $2$ for a probable LLS or DLA in the SDSS spectrum, $4$ for a BAL, and $8$ for a possible BAL. These individual flags are additive for a given object (e.g., 6 denotes a BAL with an LLS or DLA).}
\tablenotetext{c}{The observation flag is 0 if the object has not been observed with {\it HST}, 1 if it has been observed with FUV (to near 304 \AA\ rest) spectroscopy through {\it HST} Cycle 16 Supplemental, 2 if it has been observed less conclusively (NUV spectroscopy, UV imaging, or pre-COSTAR observation), and 3 if it is a planned target for {\it HST} Cycle 17 observation.}

\end{deluxetable}

\begin{deluxetable}{lrrlrccc}
\tabletypesize{\footnotesize}
\rotate
\tablewidth{550pt}
\tablecaption{Catalog of Candidate EUV-Bright (non-SDSS) VCV/{\it GALEX} Quasars
\label{tab:TargetListVeron}}

\tablehead{\colhead{Name} & \colhead{RA} & \colhead{Dec} & \colhead{Redshift} & \colhead{V\tablenotemark{a}} & \colhead{NUV flux} & \colhead{FUV flux}  & \colhead{Obs.\tablenotemark{c}} \\ 
\colhead{} & \colhead{(J2000)} & \colhead{(J2000)} & \colhead{} & \colhead{(mag)} & \colhead{($10^{-17}$)\tablenotemark{b}} & \colhead{($10^{-17}$)\tablenotemark{b}} & \colhead{} } 

\startdata
TEX0004+139 & 1.7396 & 14.2628 & 3.2 & 19.90 & 4.00 & 6.76 & 0 \\
Q0016-3936 & 4.7992 & -39.3278 & 3. & *19.79 & 22.16 & 30.82 & 0 \\
Q0016-3941 & 4.8425 & -39.4058 & 2.96 & *19.68 & 2.77 & \nodata & 0 \\
2QZJ002416-3149 & 6.0675 & -31.8289 & 2.846 & *20.24 & 2.31 & 4.36 & 0 \\
Q0023-4013 & 6.5912 & -39.9514 & 3. & *19.64 & 3.04 & \nodata & 0 \\
Q0026-3934 & 7.1929 & -39.3050 & 2.91 & *19.59 & 1.50 & \nodata & 0 \\
2QZJ002855-2937 & 7.2317 & -29.6200 & 2.785 & *19.84 & 1.85 & \nodata & 0 \\
Q0027-4132 & 7.4904 & -41.2589 & 2.79 & *19.70 & 3.89 & \nodata & 0 \\
2QZJ003130-3033 & 7.8750 & -30.5558 & 2.811 & *19.16 & 1.18 & \nodata & 0 \\
2QZJ003447-3048 & 8.6967 & -30.8033 & 2.785 & *19.97 & 5.30 & 4.23 & 0
\enddata

\tablecomments{This table is available in its entirety in a machine-readable form in the online journal. A portion is shown here for guidance regarding its form and content.}

\tablenotetext{a}{V magnitude unless otherwise marked as: photographic (*), red (R), infrared (I), or photographic O-plates (O).}
\tablenotetext{b}{erg s$^{-1}$ cm$^{-2}$ \AA$^{-1}$, from {\textit{GALEX}}}
\tablenotetext{c}{The observation flag is 0 if the object has not been observed with {\it HST}, 1 if it has been observed with FUV (to near 304 \AA\ rest) spectroscopy through {\it HST} Cycle 16 Supplemental, 2 if it has been observed less conclusively (NUV spectroscopy, UV imaging, or pre-COSTAR observation), and 3 if it is a planned target for {\it HST} Cycle 17 observation.}

\end{deluxetable}

\begin{deluxetable}{lrrlccclc}
\tabletypesize{\scriptsize}
\rotate
\tablewidth{600pt}
\tablecaption{Quasars/Sightlines with Confirming ACS Reconnaissance UV Spectra\label{tab:ACSObs}}

\tablehead{\colhead{Name} & \colhead{RA} & \colhead{Dec} & \colhead{Redshift} & \colhead{NUV flux} & \colhead{FUV flux} & \colhead{Exp. Time} & \colhead{Obs. Date} & \colhead{He II Ly$\alpha$ break} \\
\colhead{} & \colhead{(J2000)} & \colhead{(J2000)} & \colhead{} & \colhead{($10^{-17}$)\tablenotemark{a}} & \colhead{($10^{-17}$)\tablenotemark{a}} & \colhead{(s)} & \colhead{} & \colhead{} }

\startdata
SDSSJ102646.03+353145.4 & 156.691790 & 35.529284 & 3.65 & 3.04 & ... & 4225 & 2009 Apr 17 & yes\tablenotemark{b} \\
SDSSJ104255.97+512936.3 & 160.733200 & 51.493418 & 3.39 & ... & 7.32 & 4310 & 2009 Mar 9 & yes \\
SDSSJ104757.70+325023.5 & 161.990430 & 32.839871 & 4.09 & 0.83 & ... & 2895\tablenotemark{c} & 2009 Apr 17 & no \\
SDSSJ121019.97+442652.4 & 182.583190 & 44.447877 & 3.37 & 0.73 & 2.63 & 4245 & 2009 Feb 24 & yes \\ 
SDSSJ122752.02+165522.7 & 186.966740 & 16.922978 & 3.35 & 2.46 & ... & 4195 & 2009 Mar 4 & no \\
SDSSJ124306.56+530522.1 & 190.777330 & 53.089473 & 3.57 & 1.88 & 7.39 & 4310 & 2009 Apr 26 & yes \\
SDSSJ125353.71+681714.2 & 193.473810 & 68.287265 & 3.47 & 3.62 & 6.76 & 4470 & 2009 Mar 30 & yes \\
SDSSJ125903.26+621211.7 & 194.763590 & 62.203247 & 3.24 & 2.62 & 7.35 & 4395 & 2009 Mar 29 & ? \\
SDSSJ131536.57+485629.1 & 198.902390 & 48.941427 & 3.62 & ... & 5.63 & 4275 & 2009 Apr 26 & yes \\
SDSSJ133019.14+261337.7 & 202.579770 & 26.227148 & 3.24 & 1.93 & 9.33 & 4205 & 2009 Mar 11 & yes \\
SDSSJ133022.77+281333.2 & 202.594860 & 28.225881 & 3.32 & ... & 6.52 & 4205 & 2009 Mar 6 & yes \\
SDSSJ134142.04+075639.2 & 205.425180 & 7.944210 & 3.30 & 4.18 & ... & 4175 & 2009 Mar 31 & yes\tablenotemark{b} \\
SDSSJ144436.69+105549.9 & 221.152860 & 10.930520 & 3.24 & ... & 8.67 & 4185 & 2009 Apr 7 & yes \\
SDSSJ144516.47+095836.1 & 221.318610 & 9.976696 & 3.54 & 15.93 & 18.70 & 4175 & 2009 Apr 12 & ?\\
SDSSJ144623.01+434308.7 & 221.595870 & 43.719091 & 3.17 & 2.12 & 8.28 & 4245 & 2009 Apr 9 & yes \\
SDSSJ153930.24+285035.4 & 234.876020 & 28.843156 & 3.65 & 2.32 & 8.98 & 1600\tablenotemark{d} & 2009 Apr 8 & ? \\
SDSSJ155531.64+384924.9 & 238.881850 & 38.823579 & 3.50 & ... & 4.74 & 4225 & 2009 Mar 5 & yes \\
SDSSJ161426.82+485958.7 & 243.611730 & 48.999644 & 3.81 & 1.98 & 3.22 & 4275 & 2009 Apr 7 & yes \\
QJ1943.9-1502 & 295.9950 & -15.0469 & 3.3\tablenotemark{e} & 6.88 & 13.03 & 2110\tablenotemark{d} & 2009 May 3 & ? \\
\enddata

\tablenotetext{a}{erg s$^{-1}$ cm$^{-2}$ \AA$^{-1}$, from {\it GALEX}}
\tablenotetext{b}{May not be suitable for follow-up, despite the He II break, because of substantial UV absorption redward of the break as well.}
\tablenotetext{c}{This quasar had two particularly noisy or misaligned sub-exposures. Our analysis does not use these sub-exposures, nor have they been included in the tabulated exposure time.}
\tablenotetext{d}{Due to problems acquiring guide stars, there is only one orbit of {\it HST} observations for these objects.}
\tablenotetext{e}{Uncertain, as it is based on a single line.}

\end{deluxetable}

\end{document}